\documentclass[
    ,final            
  ]
  {aipproc}

\layoutstyle{6x9}

\begin{document}

\title{Correlations in superstatistical systems}


\author{Christian Beck}{
  address={School of Mathematical Sciences, Queen Mary, University of London,
Mile End Road, London E1 4NS, UK}
}



\begin{abstract}
We review some of the statistical properties of
higher-dimensional superstatistical stochastic
models. As an example, we analyse the
stochastic properties of a superstatistical model
of 3-dimensional Lagrangian turbulence, and compare with
experimental data. Excellent agreement is obtained for various
measured quantities, such as acceleration probability densities,
Lagrangian scaling exponents, 
correlations between acceleration components, and time decay of correlations.
We comment on how to proceed from superstatistics to a thermodynamic description.
\end{abstract}

\maketitle


\section{A short reminder: What is superstatistics?}\label{sec1}
%



Complex systems often exhibit a dynamics on two time scales:
A fast one as represented by a given stochastic process and a slow one
for the parameters of that process.
As a very simple example consider the following linear
Langevin equation
\begin{equation}
 \dot{v}=-\gamma v + \sigma L(t)
\end{equation}
with
parameters $\gamma , \sigma$ that fluctuate on a long
time scale \cite{prl2001}. It describes
the velocity $v$ of a
Brownian particle that moves through spatial `cells'
with different local inverse temperature $\beta:=\gamma /(2\sigma^2)$ in each cell (a
nonequilibrium situation).
Assume, for example, that the probability distribution of $\beta$ in the various
cells is a $\chi^2$-distribution of degree $n$,
\begin{equation}
f(\beta) \sim \beta^{n/2-1} e^{-\frac{n\beta}{2\beta_0}}.
\end{equation}
Then the conditional probability is $p(v|\beta))\sim
e^{-\frac{1}{2} \beta v^2}$, the joint probability is $p(v,
\beta)=f(\beta) p(v|\beta)$, and the marginal probability is
$p(v)=\int_0^\infty f(\beta) p(v|\beta) d\beta$. Integration
yields
\begin{equation}
p(v)\sim \frac{1}{(1+\frac{1}{2}\tilde{\beta}(q-1)v^2)^{1/(q-1)}}
\end{equation}
i.e.\ we obtain power-law stationary distributions 
just as in Tsallis statistics \cite{tsa1} with
$q=1+\frac{2}{n+1}$, $\tilde{\beta}=2\beta_0/(3-q)$, 
where $\beta_0=\int f(\beta) \beta d\beta $
is the average of $\beta$.

All this has a very broad interpretion and
can be generalized in various ways---$\beta$ need not to be
inverse temperature. One can generalize the above example to {
general probability densities $f(\beta)$} and {general
Hamiltonians} in statistical mechanics. One then  has a superposition of two different
statistics: that of $\beta$ and that of ordinary statistical mechanics. The
short name for this is {\em superstatistics} \cite{beck-cohen}.
Superstatistics describes complex {\em nonequilibrium systems}
with {\em spatio-temporal fluctuations of an intensive parameter}
(e.g.\ inverse temperature) on a large scale.

Define an effective Boltzmann factor $B(E)$ by
\begin{displaymath}
B(E)=\int_0^\infty f(\beta) e^{-\beta E} d\beta
\end{displaymath}
where $f(\beta)$ is the probability distribution of $\beta$
and $E$ the energy of the system. Many
results can be proved for {\em general $f(\beta)$}. Here we list
some recent theoretical developments of the superstatistics
concept:
\begin{itemize}

\item Can prove superstatistical generalizations of fluctuation
theorems \cite{supergen}

\item Can develop a variational principle for the large-energy
asymptotics of general superstatistics \cite{touchette-beck}
(depending on $f(\beta)$, one can get not only power laws for
large $E$ but e.g. also stretched exponentials)

\item Can formally define generalized entropies for general
superstatistics \cite{souza, abc}

\item Can study various theoretical extensions and workouts of the
superstatistics concept \cite{chavanis, vignat, crooks,
naudts, rodri, luba}

\item Can prove a superstatistical version of a Central Limit
Theorem leading to Tsallis statistics \cite{vignat2}

\item Can relate it to fractional reaction equations
\cite{haubold}

\item Can consider superstatistical random matrix theory
\cite{abul-magd}

\item Can apply superstatistical techniques to networks
\cite{abe-thurner} and time series \cite{BCS}

\end{itemize}

...and some more practical applications:

\begin{itemize}

\item Can apply superstatistical methods to analyse the statistics of 3d
hydrodynamic turbulence \cite{prl2001, BCS, prl2007, reynolds, euro,
boden2}

\item Can apply it to atmospheric turbulence (wind velocity
fluctuations at Florence airport \cite{rapisarda, rap2})
and defect turbulence \cite{daniels}

\item Can apply superstatistical methods to finance \cite{bouchard, ausloos}

\item Can apply it to solar flares \cite{maya},
 and even to print queues
\cite{maya2}

\item Can apply it to cosmic ray statistics \cite{cosmic}

\item Can apply it to various scattering processes in particle
physics \cite{wilk, wilk2}

\item Can apply it to hydroclimatic fluctuations \cite{porporato}

\item Can apply it to British train delay statistics \cite
{briggs}

\end{itemize}






















\section{Physically relevant superstatistical universality classes}\label{sec2}

Basically, there are {\em 3 physically relevant universality
classes} \cite{BCS}:

\begin{itemize}
\item (a) $\chi^2$-superstatistics ($=$ Tsallis statistics)
\item (b) inverse $\chi^2$-superstatistics
\item (c) lognormal superstatistics
\end{itemize}

Why? Consider, e.g., case (a). Assume there are many microscopic
random variables $\xi_j$, $j=1,\ldots , J$, contributing to
$\beta$ in an additive way. For large $J$, their sum
$\frac{1}{\sqrt{J}}\sum_{j=1}^J\xi_j$ will approach a Gaussian
random variable $X_1$ due to the (ordinary) Central Limit Theorem. There can
be $n$ Gaussian random variables $X_1,\ldots ,X_n$ due to various
relevant degrees of freedom in the complex system. Since $\beta$ is positive
we may square the $X_i$ to obtain something positive. The sum 
$\beta=\sum_{i=1}^nX_i^2$ is then $\chi^2$-distributed
with degree $n$, i.e.,
\begin{equation}
f(\beta )=\frac 1{\Gamma (\frac n2)}\left( \frac n{2\beta
_0}\right) ^{n/2}\beta ^{n/2-1}e^{-\frac{n\beta }{2\beta _0}},
\label{chi2}
\end{equation}
where $\beta_0$ is the average of $\beta$.
Integration as described in section 1 yields Tsallis statistics as a special
case of superstatistics.


(b) The same considerations can be applied if the `temperature'
$\beta^{-1}$ rather than $\beta$ itself is the sum of several
squared Gaussian random variables arising out of many microscopic
degrees of freedom $\xi_j$. The resulting $f(\beta)$ is the
inverse $\chi^2$-distribution:
\begin{equation}
f(\beta )=\frac{\beta _0}{\Gamma (\frac n2)}\left( \frac{n\beta
_0}2\right) ^{n/2}\beta ^{-n/2-2}e^{-\frac{n\beta _0}{2\beta }}.
\label{chi2inv}
\end{equation}
It generates superstatistical distributions $p(E)\sim \int
f(\beta) e^{-\beta E}$ that decay as
$e^{-\tilde{\beta}\sqrt{E}}$ for large $E$ \cite{touchette-beck}.


(c) $\beta$ may be generated by multiplicative random processes.
Consider a local cascade random variable $X_1= \prod_{j=1}^{J}
\xi_j$, where $J$ is the number of cascade steps and the $\xi_j$
are positive microscopic random variables. By the Central Limit
Theorem, $\frac{1}{\sqrt{J}} \log X_1=
\frac{1}{\sqrt{J}} \sum_{j=1}^J \log \xi_j$ becomes Gaussian for
large $J$. Hence $X_1$ is log-normally distributed. In general
there may be $n$ such product contributions to $\beta$, i.e.,
$\beta = \prod_{i=1}^n X_i$. Then $\log \beta = \sum_{i=1}^n \log
X_i$ is a sum of Gaussian random variables; hence it is Gaussian
as well. Thus $\beta$ is log-normally distributed, i.e.,
\begin{equation}
f(\beta )=\frac{1}{\sqrt{2\pi}s\beta} \exp \left\{ \frac{-(\ln
\frac{\beta}{m})^2}{2s^2}\right\}. \label{logno}
\end{equation}
Lognormal superstatistics is relevant in turbulence \cite{BCS, prl2007, reynolds, euro, boden2}.



\section{Application to Lagrangian turbulence}\label{sec3}


Turbulence is a spatio-temporal chaotic state of the Navier-Stokes
equation. Energy is dissipated in a cascade-like process.
Bodenschatz et al. \cite{boden2, boden1, boden3} obtained rather precise
measurements of the acceleration $\vec{a}(t)$ of a single tracer
particle in a turbulent flow.
One can now construct a superstatistical Lagrangian model for {\em
3-dimensional} velocity differences 
$\vec{u}(t):=\vec{v}(t+\tau) -\vec{v} (t)$ 
of such  a tracer particle
(note
that $\vec{a}=\vec{u}/\tau$ for small $\tau$). This model is
given by the superstatistical stochastic differential equation \cite{prl2007}
\begin{equation}
\dot{\vec{u}}=-\gamma \vec{u}+B \vec{n} \times \vec{u}+\sigma
\vec{L}(t). \label{2}
\end{equation}
The new thing as compared to previous work is the term involving
the vector product. It describes fluctuating enstrophy
(rotational energy) around the test particle.
While $\gamma$ and $B$ are constants, the noise strength $\sigma$
and the unit vector $\vec{n}$ evolve stochastically on a large
time scale $T_\sigma$ and $T_{\vec{n}}$, respectively. One has
$T_\sigma \gamma \sim R_\lambda
>>1$, where $R_\lambda$ is the Taylor scale Reynolds number.
The time scale $T_{\vec{n}}$ describes the average
life time of a region of given vorticity surrounding the test
particle.

Define $\beta:=2\gamma/\sigma^2$, then in this model $ \beta^{-1}
\sim \nu^{1/2} \langle \epsilon \rangle^{-1/2} \epsilon$, where
$\nu$ is the kinematic viscosity and $\langle \epsilon \rangle$
the average energy dissipation. The probability density of the
stochastic process $\beta(t)$ is assumed to be a lognormal
distribution as given in eq.~(\ref{logno}).
For very small $\tau$ an acceleration component of the particle is given
by $a_x=u_x/\tau$ and one gets the following prediction for the stationary
distribution:
\begin{equation}
p(a_x) = \frac{\tau}{2\pi s }\int_0^\infty d\beta \; \beta^{-1/2}
\exp\left\{ \frac{-(\log \frac{\beta}{m})^2}{2s^2}\right\}
e^{-\frac{1}{2}\beta \tau^2 a_x^2}  \label{10}
\end{equation}
This compares very well with the experimentally measured
probability distribution of acceleration, see Fig.~1.
\begin{figure}
\includegraphics[width=10cm]{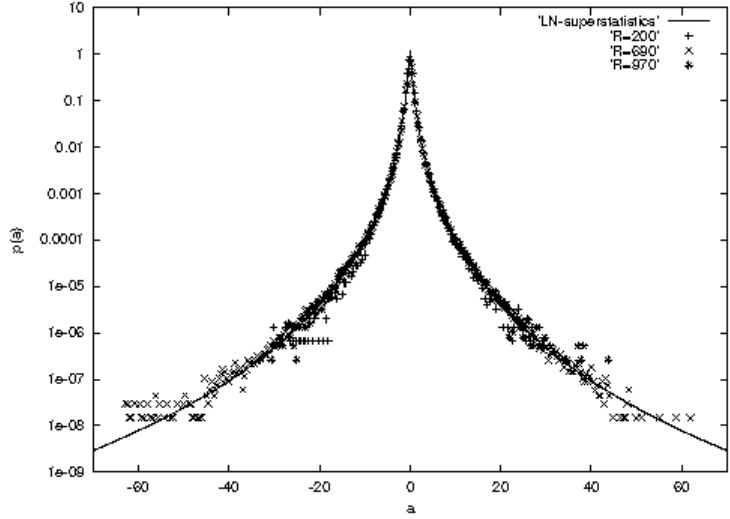}
\caption{Distribution of acceleration as measured by Bodenschatz
\cite{boden2,boden1} and as predicted by eq.~(\ref{10}), $s^2\approx 3$
.}
\end{figure}

\section{Correlations induced by superstatistics}\label{sec4}
3-dimensional superstatistics induces correlations between the
components $a_x,a_y,a_z$ of the acceleration vector $\vec{a}$. Consider the ratio
$R:=p(a_x,a_y)/(p(a_x)p(a_y))$. For independent acceleration
components this ratio would always be given by $R=1$. However,
our 3-d superstatistical model yields the prediction
\begin{equation}
R=\frac{\int_0^\infty \beta f(\beta)e^{-\frac{1}{2}\beta \tau^2
(a_x^2+a_y^2)}d\beta}{ \int_0^\infty\beta^{1/2}f(\beta
)e^{-\frac{1}{2}\beta \tau^2 a_x^2}d\beta
\int_0^\infty\beta^{1/2}f(\beta)e^{-\frac{1}{2}\beta \tau^2
a_y^2}d\beta}. \label{rrr}
\end{equation}
This is a very general formula, it is also valid for Tsallis
statistics, where $f(\beta)$ is the $\chi^2$-distribution. Note
that $R=1$ for $f(\beta)=\delta (\beta -\beta_0)$, i.e.\
if there are no fluctuations in $\beta$ then all components
are independent random variables.

Fig.~2 shows $R=p(a_x,a_y)/(p(a_x)p(a_y))$ as predicted by
lognormal superstatistics.
\begin{figure}
\includegraphics[width=10cm]{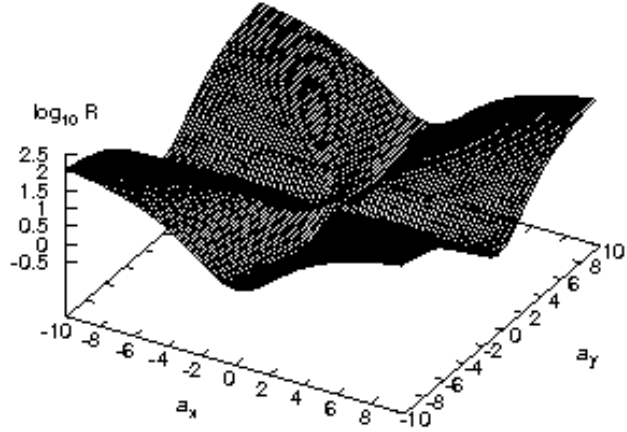}
\caption{The quantitity $R$ as given by eq.~(\ref{rrr})
for lognormal superstatistics.}
\end{figure}
The figure strongly resembles $R$ as experimentally measured by
Bodenschatz et al. \cite{boden2} in a turbulent flow, see Fig.~3.

\begin{figure}
\includegraphics[width=10cm]{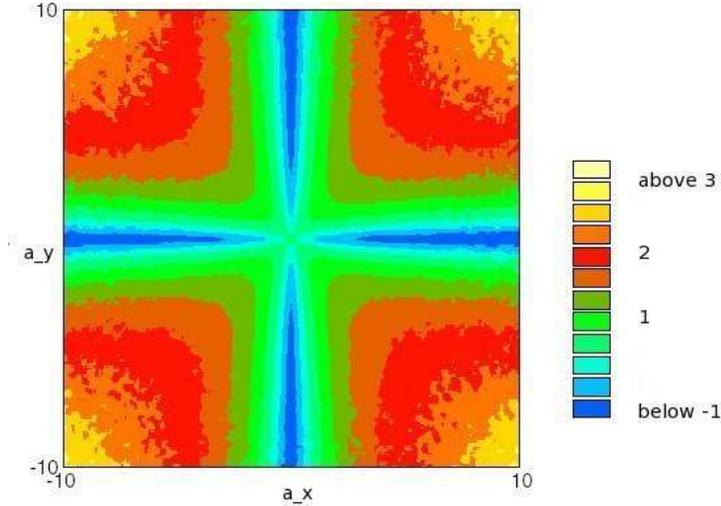}
\caption{$\log_{10} R$ as measured by Bodenschatz et al. \cite{boden2}.}
\end{figure}

Besides correlations between components one can also look at
temporal correlations. The superstatistical model \cite{prl2007} allows for the
calculation of temporal correlation functions as well. In
particular, we may be interested in temporal correlation functions
of single components $u_x$ of velocity differences, i.e.\ $C(t)=
\langle u_x(t'+t) u_x(t') \rangle$. By averaging over the
possible random vectors $\vec{n}$ one arrives at the formula
\begin{equation}
C(t)=\frac{1}{3} \langle u_x^2 \rangle e^{-\gamma t} (2\cos
 B t +1),
\end{equation}
i.e.\ there is rapid (exponential) decay with a zero-crossing at
$t^*=\frac{2}{3}\pi B^{-1}$. Exponential decay and zero-crossings
are also observed for the experimental data. The model \cite{prl2007} also
correctly reproduces the experimentally observed fact that the
correlation function of the absolute value $|\vec{a}|$ decays
very slowly as compared to that of the single components. Moreover, it correctly
describes the fact that
enstrophy lags behind dissipation \cite{zeff}.

\section{Lagrangian scaling exponents} \label{sec5}
The moments of velocity differences of a single Lagrangian test
particle that is embedded in a turbulent flow scale differently
from those measured in a fixed laboratory frame. Our
superstatistical model \cite{prl2007}  allows for the analytic evaluation of the
Lagrangian scaling exponents. The moments of velocity difference
components $u_x$ on a time scale $\tau$ are obtained as
\begin{equation}
\langle u_x^j \rangle =
(j-1)!! m^{-\frac{j}{2}} w^{\frac{1}{8}j^2}. \label{super}
\end{equation}
Assuming simple scaling laws of the form $m\sim \tau^a$,
$w=e^{s^2} \sim \tau^b$, where $a$ and $b$ are so far arbitrary
real numbers, on gets {\em $\langle u_x^j\rangle \sim
\tau^{\zeta_j}$} $\sim \tau^{-a\frac{j}{2}+b\frac{1}{8}j^2} $.
Hence the Lagrangian scaling exponents are given by
$\zeta_j=-\frac{a}{2}j+\frac{b}{8}j^2$.
Usually one assumes $\zeta_2=1$, hence we get $a=\frac{1}{2}b-1$
thus
\begin{equation}
\zeta_j=(\frac{1}{2}+\lambda^2)j-\frac{1}{2}\lambda^2 j^2,
\label{scalex}
\end{equation}
where $\lambda^2:=-\frac{1}{4}b$.
This prediction is in good agreement with the recent measurements
of Bodenschatz et al. \cite{boden3}, see Fig.~4.
\begin{figure}
\includegraphics[width=10cm]{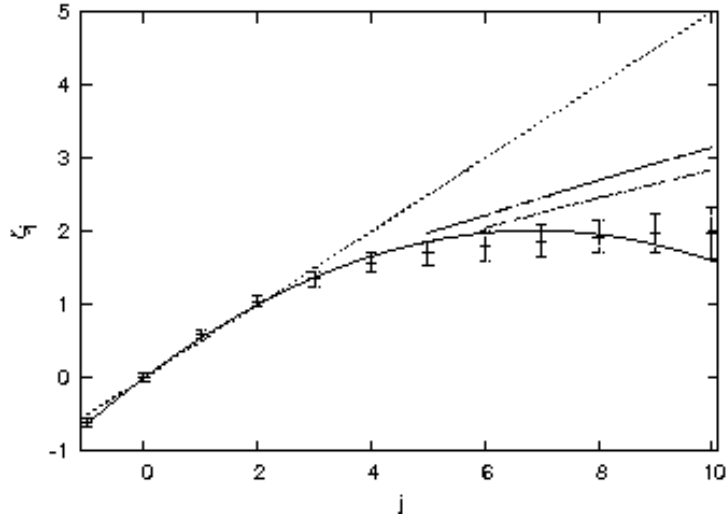}
\caption{Lagrangian scaling exponents. Data points: Measurements
of Bodenschatz et al. \cite{boden3}. Solid line: Theoretical
prediction of the superstatistical model ($\lambda^2=0.085$)
\cite{prl2007}. Dashed lines: Some other competing models
\cite{chev}.}
\end{figure}

\section{From superstatistics to (generalized) thermodynamics}\label{sec6}
We end this paper with some more general thoughts. Can we proceed
from superstatistics as a merely statistical technique to a proper
{\em thermodynamic description}? There are some early attempts in
this direction by Tsallis and Souza \cite{souza}. Here we want
to follow a somewhat different approach
\cite{abc}: One starts quite generally from two random variables
$E$ and $B$ (representing energy and invere temperature) and then
considers the following effective entropy for a
superstatistical system
\begin{eqnarray*}
S[E,B]&=&S[E|B]+S[B] \\
&=&\int d\beta f(\beta) (\beta U(\beta)+\ln Z(\beta))- \int d\beta
f( \beta) \ln f(\beta),
\end{eqnarray*}
where $U$ is the local internal energy and $Z$ the local partition function.
One can do thermodynamics with this extended entropy
function. It reduces to ordinary thermodynamics for
$f(\beta)=\delta (\beta -\beta_0)$.
For sharply peaked distributions $f(\beta)$ this is a
slightly deformed thermodynamics, which can be evaluated in a
perturbative way.
One can also maximize this entropy with respect to appropriate
constraints {in $\beta$} to get e.g. a lognormal distribution
for $f(\beta)$, or generally some other distribution $f(\beta)$
depending on the constraints. For more details, see \cite{abc}.

\section{Summary}

\begin{itemize}

\item
{Superstatistics} (a `statistics of a statistics') provides a
physical reason why more general types of Boltzmann factors 
(e.g.\ of power-law type) are relevant for {nonequilibrium} systems
with fluctuations of an intensive parameter.

\item
There is evidence for three major physically relevant {
universality classes}: $\chi^2$-superstatistics $=$ Tsallis
statistics, inverse $\chi^2$-superstatistics, and lognormal
superstatistics. These arise as {universal limit statistics} for
many different complex systems.

\item
Superstatistical techniques have been successfully applied to a {
variety of complex systems.}

\item
A superstatistical model of {Lagrangian turbulence} \cite{prl2007}
is in excellent agreement with the experimental data for
probability densities, correlations between components, decay of
correlations, and Lagrangian scaling exponents.

\item
The long-term aim is to find a good {thermodynamic description}
for general superstatistical systems.

\end{itemize}

\end{document}